\documentclass[showpacs,amsmath,amssymb,aps,showkeys,floatfix,prd,a4paper]{revtex4}

\usepackage{dcolumn}
\usepackage{bm}
\usepackage{epsfig}
\usepackage{amsfonts}
\usepackage{amssymb,amscd}
\usepackage{xcolor}

\def\lsim{\raise0.3ex\hbox{$<$\kern-0.75em\raise-1.1ex\hbox{$\sim$}}}

\def\gsim{\raise0.3ex\hbox{$>$\kern-0.75em\raise-1.1ex\hbox{$\sim$}}}

\newcommand{\be}{\begin{equation}}

\newcommand{\ee}{\end{equation}}

\def\beq{\begin{equation}}

\def\eeq{\end{equation}}

\def\beqa{\begin{eqnarray}}

\def\eeqa{\end{eqnarray}}

\newcommand{\ba}{\begin{eqnarray}}

\newcommand{\ea}{\end{eqnarray}}

\def\gappeq{\mathrel{\rlap {\raise.5ex\hbox{$>$}}

{\lower.5ex\hbox{$\sim$}}}}

\def\lappeq{\mathrel{\rlap{\raise.5ex\hbox{$<$}}

{\lower.5ex\hbox{$\sim$}}}}

\def\Toprel#1\over#2{\mathrel{\mathop{#2}\limits^{#1}}}

\begin{document}

%\title{Investigating the impact of the modeling of Earth structure \\ on the neutrino propagation at ultra - high - energies}

\title{Sensitivity of the neutrino transmission coefficient at high energies to the Earth's density profile}

\author{Reinaldo {\sc Francener}}
\email[(Corresponding author) Electronic address: ]{reinaldofrancener@gmail.com}
\affiliation{Instituto de Física Gleb Wataghin - UNICAMP, 13083-859, Campinas, SP, Brazil. }

\author{Victor P. {\sc Gon\c{c}alves}}
\email{barros@ufpel.edu.br}
\affiliation{Institute of Physics and Mathematics, Federal University of Pelotas, \\
  Postal Code 354,  96010-900, Pelotas, RS, Brazil}
\affiliation{Institute of Modern Physics, Chinese Academy of Sciences,
  Lanzhou 730000, China}

\author{Diego R. {\sc Gratieri}}
\email{drgratieri@id.uff.br}
\affiliation{Escola de Engenharia Industrial Metal\'urgica de Volta Redonda,
Universidade Federal Fluminense (UFF),\\
 CEP 27255-125, Volta Redonda, RJ, Brazil}
\affiliation{Instituto de Física Gleb Wataghin - UNICAMP, 13083-859, Campinas, SP, Brazil. }

\begin{abstract}
The flux of atmospheric and astrophysical neutrinos measured  in ultra - high - energy neutrino detectors  is  strongly dependent  on the description of the propagation and absorption of the neutrinos during the passage through Earth to the detector. In particular, the attenuation of the incident neutrino flux depends on the details of the matter structure between the source and the detector.  In this paper, we will investigate the impact of different descriptions for the density profile of Earth on the transmission coefficient, defined as the ratio between the  flux measured in the detector and the incoming neutrino flux. We  will consider five different models for the Earth's density profile and estimate how these different models modify the target column density and  transmission coefficients for different flavors. The results are derived by solving the cascade equations taking into account the  neutral current interactions and tau regeneration. A comparison with approximated solutions is also presented. Our results indicated that the predictions are sensitive to the model considered for the density profile, with the simplified three layer model providing a satisfactory description when compared with the Preliminary Reference Earth Model results. Our results also showed that the more simplified models with two or one layer fail mainly for neutrinos that cross a small column of matter and are not good approximations for the neutrino propagation. These findings highlight the importance of using realistic Earth density models in ultra-high-energy neutrino analyses, as simplified models may lead to misestimations in the predicted attenuation effects.
 
\end{abstract}

%\pacs{12.38.-t, 24.85.+p, 25.30.-c}


\maketitle

%\vspace{1cm}

\section{Introduction}

\begin{figure}[t]
	\centering
	\includegraphics[width=0.45\textwidth]{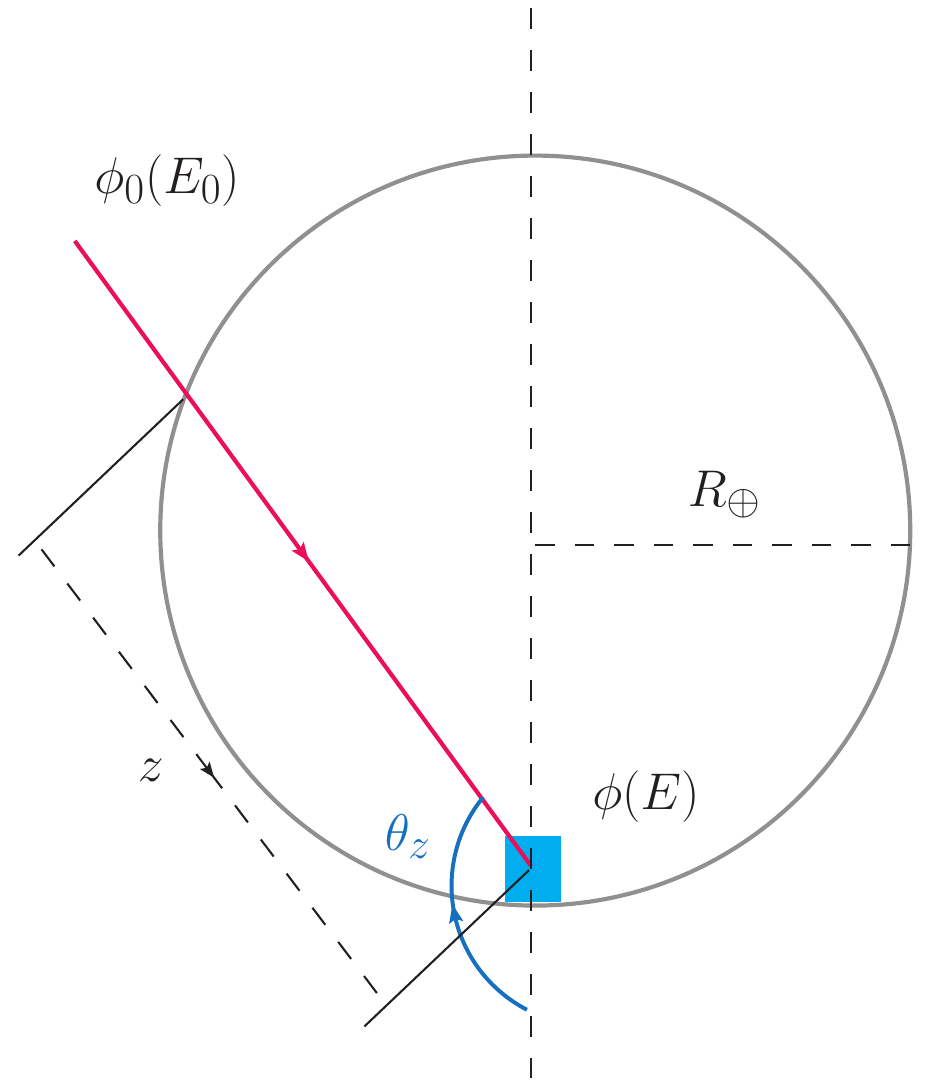}
\caption{Neutrino path towards an underground detector. The flux of neutrinos with energy $E$ at the detector ($\phi (E)$) is determined by the incoming neutrino flux ($\phi_0 (E_0)$), the distance traveled through the Earth $z$, the density profile of the Earth and the CC and NC neutrino - nucleon cross - sections. }
\label{fig:path}
\end{figure}

One of the main goals of the current and future neutrino observatories is the study of ultra - high - energy (UHE) neutrino events, which are expected to improve our understanding about the origin, propagation, and interaction of neutrinos. Over the last years, the IceCube Neutrino Observatory  has detected several events with deposited energies in the range between TeV and PeV,  produced  by cosmic rays in the atmosphere and by astrophysical sources (For  recent reviews see, e.g., Refs. \cite{Ackermann:2022rqc,MammenAbraham:2022xoc,Reno:2023sdm}). The interpretation of these measurements for the neutrino flux with a given energy $E$ at the detector ($\phi (E)$) depends on the precise knowledge of the incoming neutrino flux  ($\phi_0 (E_0)$, with $E_0 > E$), the amount of matter traveled (which is determined by the zenith angle $\theta_z$), the model for the density profile of the Earth and the neutrino - matter interaction cross-sections (See Fig. \ref{fig:path}). As these events probe an unexplored regime of neutrino - nucleon interactions, characterized by very small values for the Bjorken variable $x$ and large values for the gauge boson virtuality $Q^2$,  many theoretical studies were performed during the last decade (See, e.g. Refs. \cite{CS,Goncalves:2010ay,Illarionov:2011wc,Block:2013nia,Goncalves:2015fua,Albacete:2015zra,Arguelles:2015wba,bgr18,Goncalves:2021gcu,Valera:2022ylt,Goncalves:2022uwy}),  mainly focused on an improved description of the  neutrino - nucleon cross-section at high energies, where new non-linear effects, associated with the high partonic density within the nucleons, are expected to modify the dynamics of the strong interactions \cite{hdqcd}.  Such effects, if present, also have a direct impact on the neutrino attenuation when crossing the Earth. In addition, the impact of the modeling of the incoming neutrino flux has also been estimated in recent studies \cite{Goncalves:2022uwy,Fiorillo:2022rft}. A less exploited topic, is the modeling of the Earth structure { on the attenuation of UHE neutrinos}, which is the focus of this study. 

The description of the Earth structure is mainly based on seismic studies \cite{Dziewonski:1981xy,bolt,kennett,masters,wit}, where the propagation of seismic waves inside the Earth reveals the properties of matter. As the neutrino propagation within the Earth does depend on the details of the matter structure between the source and the detector, the study of the neutrino absorption when passing through the Earth offers an opportunity to infer its profile density. Such an alternative was proposed many years ago \cite{Volkova:1974xa,DeRujula:1983ya,Wilson:1983an} and discussed in detail in Refs. \cite{Jain:1999kp,Reynoso:2004dt,Gonzalez-Garcia:2007wfs,Borriello:2009ad,Denton:2020jft,Kumar:2021faw,Hajjar:2023knk}. Recent results \cite{Donini:2018tsg} derived using the one - year sample of thoroughgoing muons produced by atmospheric muon neutrinos, have demonstrated that the Earth tomography with neutrinos is feasible.  
Moreover, it is a well known fact that the most important matter effect on the neutrino propagation while they cross the Earth takes place on the neutrino oscillation probabilities (For a recent study see e.g. Ref. \cite{Denton:2021rgt}). Indeed, due to the interaction with the electrons in the medium, it is possible to occur resonant oscillatory behavior through the Mikheyev - Smirnov - Wolfenstein (MSW) effect \cite{Wolfenstein,MS}, and variations of the Earth's density profile would lead to modifications in both neutrino energy and incoming neutrino direction for which the resonance occurs. Even apart from the resonance region, neutrino oscillations are largely impacted by matter effects, and this phenomenon can be used to determine the Earth's density profile. See \cite{Kelly:2021jfs} for a pedagogical review on neutrino Earth's Tomography. However, as it is shown for instance in Fig. 3 of \cite{Gratieri:2017anb}, it is clear that neutrino oscillations are more relevant at $E_{\nu}\lessapprox 100\,\mathrm{GeV}$ (straightforward calculations shows that for the upcoming neutrino direction, $\mathrm{cos}\,\theta_{z} = -1.0$, the most energetic minimum on the muon neutrino oscillation probability occurs at $E_{\nu}\approx 24.5\,\mathrm{GeV}$, while the resonance due to the larger neutrino mass difference is for $E_{\nu}\approx 5$ GeV, as it is clear from Fig. 3 of \cite{Gratieri:2017anb}.). This is because the oscillatory terms depend on $\Delta m^{2}_{ij} L/E_{\nu}$, where $L$ is the distance traveled by neutrinos, and $\Delta m^{2} = m^{2}_{j}-m^{2}_{i}$ is the squared mass difference of the neutrino mass eigenstates $i$ and $j$. Henceforth, as in this work we are interested in the High Energy Start Events (HESE) neutrinos, where visible energy $\geq 60\,\mathrm{TeV}$, we can safely neglect the neutrino oscillation effects. 
Our goal in this paper is not to perform a tomographic study of the Earth assuming a particular  neutrino source and a given set of experimental data, but instead to estimate the uncertainty on the predictions for the neutrino absorption associated with the modeling of the Earth structure. In our analysis, we will assume five different models for the  density profile, and we will  compute the attenuation of high - energy neutrinos during their passage through the Earth towards large - volume detectors such as IceCube and KM3Net. In particular, we will solve, for each of these density profiles, the cascade equations for the different neutrino flavors taking into account of the neutrino attenuation and regeneration and present predictions for the energy and zenith angle dependencies of the transmission coefficient $T \equiv \phi (E)/\phi_0 (E_0)$ (See Fig. \ref{fig:path}). Finally, the difference between these predictions and those derived using the Preliminary Reference Earth Model (PREM) \cite{Dziewonski:1981xy} will be estimated. 

This paper is organized as follows.
In the next section, we review the cascade equations that govern the neutrino attenuation and regeneration.  Moreover, we present the models considered in our analysis for the Earth density profile. 
In Section \ref{sec:results} we present our predictions for the transmission coefficient considering the distinct models for the profile of Earth and estimate the difference between these results with those using the PREM profile. Finally, in Section \ref{sec:conc}, we summarize our main results and conclusions.

\section{Formalism}
In order to estimate the absorption of neutrinos as  they propagate through the Earth, we should to take into account that  charged or neutral current interactions can occur \cite{book}. In addition, for the case of electron antineutrinos, the contribution of the Glashow resonance at $E_\nu \approx 6.3$ PeV becomes non - negligible and must be included in the calculations. One has that  for electron and muon neutrinos, the CC interactions lead to the production of the associated charged lepton and the consequent attenuation of the neutrino flux. On the other hand, the NC one results in a degradation of the neutrino energy. In contrast, for tau neutrinos we should take into account of the regeneration process associated with the tau production in CC interactions that decay back to tau neutrinos, enhancing the neutrino flux at lower energies  \cite{MammenAbraham:2022xoc}. All these possibilities should be included in the cascade equations that describe the change of the neutrino flux as it traverses the Earth. In recent years, several groups have developed simulation codes that solve these equations and compute the attenuation of high - energy neutrinos \cite{Niess:2018opy,Safa:2021ghs,Garg:2022ugd}. In the current analysis, instead of use one of them, we have developed our own code, which allow us to implement in an easier way the different density profiles for the Earth. It is important to emphasize that we have verified that our results for the transmission coefficient are similar to those derived using the NuPropEarth software package and the PREM profile, which were presented in Ref.   \cite{Garcia:2020jwr}.  In what follows, we will brief review of the cascade equations that are solved to derive our predictions. For a more detailed discussion, we refer the interested reader to the Refs. \cite{edo1,edo2,edo3,edo4,edo5,edo6,edo7,edo8,Alvarez-Muniz:2018owm}.

As discussed above, in the case of tau neutrinos, one has that take into account that the tau  generated in a CC interaction can have a decay length smaller than the mean free path in the Earth, implying in its decay back into a $\nu_{\tau}$ with smaller energy. As a consequence, in order to derive realistic predictions for the tau - neutrino flux, one should to solve a system of coupled equations, given by \cite{edo7,edo8}
\begin{equation}
\begin{aligned}
\dfrac{\partial \phi_{\nu_{\tau}} (E , X)}{\partial X}  = 
-N_A [\sigma^{CC}_{\nu_{\tau}N} (E) + \sigma^{NC}_{\nu_{\tau}N}(E)] \phi_{\nu_{\tau}} (E, X) + 
N_A \int_{0}^{1}\dfrac{\mathrm{d}y}{1-y}
\dfrac{\mathrm{d} \sigma^{NC}(E / (1-y), y)}{\mathrm{d}y} 
\phi_{\nu_{\tau}} (E /(1-y) , X) + \\ 
+ \frac{1}{(E/m_{\tau})T_\tau \rho (X)}\int^{1}_{0}\mathrm{d}y\frac{\mathrm{d}n_{\tau \rightarrow \nu_{\tau}}(1-y)}{\mathrm{d}y}\phi_\tau(E/(1-y),X)
\end{aligned}
\label{eq:fluxoNu}
\end{equation}
and
\begin{eqnarray}
\begin{aligned}
\dfrac{\partial \phi_\tau (E , X)}{\partial X} = 
- \frac{1}{(E/m_{\tau})\tau \rho (X)} \phi_\tau(E, X)
+N_A \int^{1}_{0}\frac{\mathrm{d}y}{1-y}\dfrac{\mathrm{d} \sigma^{CC}(E / (1-y), y)}{\mathrm{d}y}\phi_{\nu_{\tau}}(E/(1-y),X) \,\,,
\label{eq:fluxoTau}
\end{aligned}
\end{eqnarray}
where $N_{A}$ is the Avogadro's number, $m_\tau$ and $T_\tau$  are the tau mass and lifetime, and $\rho (X)$ is the Earth density profile. Moreover,  $\phi_{\nu_\tau} (E , X)$ and $\phi_{\tau} (E , X)$ are, respectively, the differential energy spectrum of neutrino and tau at a column depth $X$ defined by
\begin{eqnarray}
    X(\theta_z) = \int_{0}^{r(\theta_z)}\rho(r)\mathrm{d}r\, ,
    \label{eq:column}
\end{eqnarray}
with $r(\theta_{z}) = -2 \,R_{\oplus} \cos\theta_{z}$ being the total distance travelled by neutrinos ($R_{\oplus}$ is the Earth radius). The distribution $\mathrm{d}n_{\tau \rightarrow \nu_\tau}(y)/\mathrm{d}y$  is calculated using the parametrization provided in Refs.  \cite{paramet1,Garg:2022ugd}.
The differential cross-section on the inelasticity $y$ for a charged current deep inelastic tau - neutrino  - nucleon interaction is given by 
\begin{eqnarray}
\frac{\mathrm{d}\sigma^{\nu_{\tau}N}}{\mathrm{d}y}  =  \int \frac{\mathrm{d}\sigma^{\nu_{\tau}N}}{\mathrm{d}x\mathrm{d}y} \mathrm{d}x\, ,
\end{eqnarray}
where the double differential cross-section is given by \cite{paschos,reno}
\begin{eqnarray}
\begin{aligned}
\frac{\mathrm{d}\sigma^{\nu_{\tau}N}}{\mathrm{d}x\mathrm{d}y}=
\frac{G_F^{2}m_N E}{\pi}
\left(
\frac{M_W^2}{Q^2+M_W^2}
\right)^2
\left\{
\left(
y^2x+\frac{m_\tau^2y}{2E m_N}
\right)F^{CC}_1(x,Q^2)+\right. 
\left(
1-y-\frac{m_\tau^2}{4E^2}-\frac{m_Nxy}{2E}
\right)F^{CC}_2(x,Q^2)+ \\
 + \left(
xy-\frac{xy^2}{2}-\frac{m_\tau^2y}{4E m_N}
\right)F^{CC}_3(x,Q^2)+ 
\left. \frac{m_\tau^{2} (m_\tau^{2}+Q^2)}{4E^2 m_N^2 x}F^{CC}_4(x,Q^2)-\frac{m_\tau^2}{E m_N}F^{CC}_5(x,Q^2)
\right\} \, ,
\end{aligned}
\end{eqnarray}
with $x$ being the Bjorken variable, $G_{F}$  the Fermi's constant,  $M_{W}$ the  $W$ boson mass, $Q^2$ the gauge boson virtuality and $F^{CC}_{i}$ are the CC nucleon structure functions, which can be expressed in terms of the parton distribution functions (PDFs) of the nucleon.   The corresponding expression for the case of an incoming antineutrino can be obtained by reversing the sign of the $F_3$ term. A similar formula can be derived for NC scattering (see, e.g., Ref. \cite{reno}).  The cascade equations for the electron and muon (anti) neutrinos differ from  Eq. (\ref{eq:fluxoNu}) by the  distribution $\mathrm{d}n/\mathrm{d}y$ in  the third term. Finally, for the particular case of the electron antineutrino, in order to take into account of the Glashow resonance, we have added the $\bar{\nu_e}$ - electron cross-section to both the first two terms of  Eq. (\ref{eq:fluxoNu}).

%\begin{figure}[t]
%	\centering
%	\begin{tabular}{ccc}
%	\includegraphics[width=0.35\textwidth]{Fig_Earth.pdf} & \,\,\,\,\, & \includegraphics[width=0.45\textwidth]{density_-1.eps} \\
%	(a) & \, &  (b) 
%			\end{tabular}
%\caption{(a) Earth profile model characterized by  three layers, associated with the crust, mantle and core. (b) Density of the Earth as a function of the distance crossed by the neutrino considering $\mathrm{cos}\, \theta_z = -1$. }
%\label{fig:earthProfile}
%\end{figure}

\begin{figure}[t]
	\centering
	\begin{tabular}{ccc}
	\includegraphics[width=0.45\textwidth]{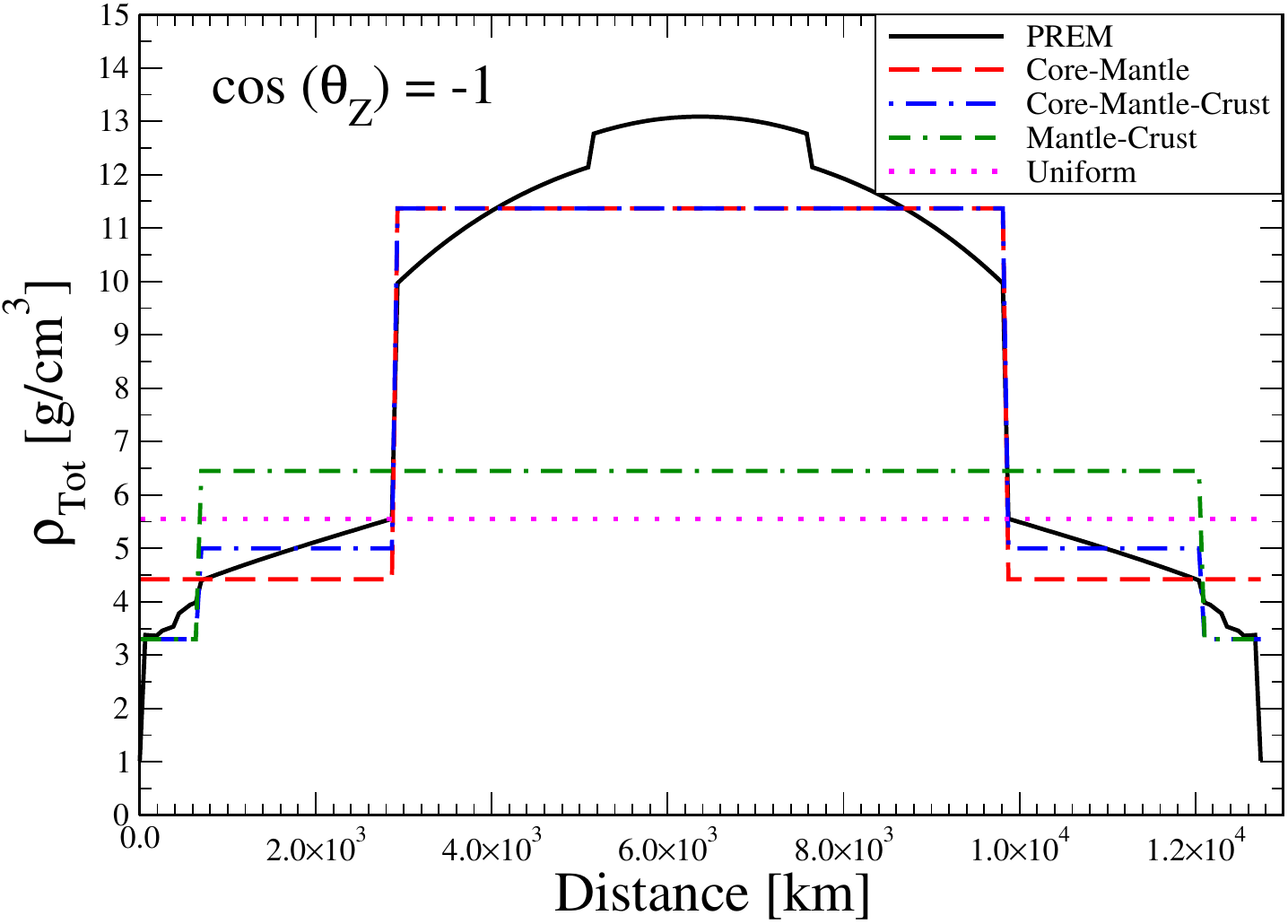} & \,\,\,\,\, & \includegraphics[width=0.45\textwidth]{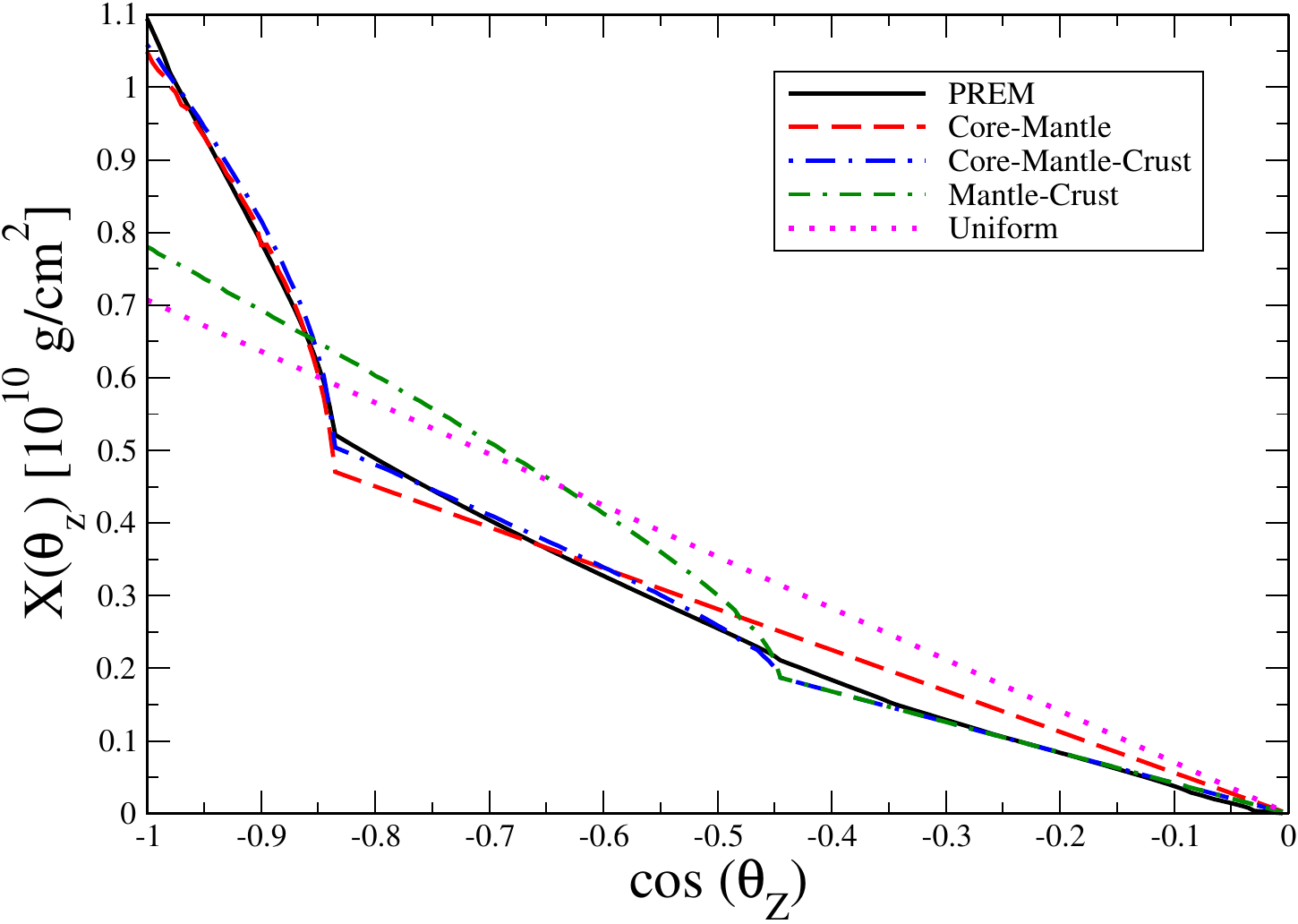} \\
	(a) & \, &  (b) 
			\end{tabular}
\caption{(a) Density of the Earth as a function of the distance crossed by an incoming  neutrino with $\mathrm{cos}\, \theta_z = -1$. (b) The target column density $X$, defined in Eq. (\ref{eq:column}),  as a function of $\mathrm{cos}\, \theta_z$ for the different models for the Earth's  density profile. }
\label{fig:earthProfile}
\end{figure}

The cascade equations provide the attenuation of  neutrinos of flavor $l$ during the passage through the Earth. In order to derive its solutions, we must specify the incoming neutrino flux $\phi^l_0(E_0) = \phi_l(E_0,X=0)$, the density profile of the Earth $\rho(r)$ and the CC and NC nucleon structure functions $F_i(x,Q^2)$. In our analysis, we will assume that the incoming neutrino flux is isotropic and  the same for all neutrino flavors and can be described by a power - like behavior, $\phi_0 \propto E_0^{-\gamma}$, where $\gamma$ is the spectral index. { In our analysis, we will assume $\gamma = 3$, but the impact of a different value will be investigated}. Moreover, the CC and NC structure functions will be estimated following Ref. \cite{reno}, assuming an isoscalar target and  using the CT14 parametrization \cite{ct14} for the proton PDFs.   The dependence of the results for the attenuation on these choices have been discussed in detail in Refs. \cite{Goncalves:2015fua,Goncalves:2021gcu}. Here, we will focus on the modeling of the density profile for the Earth. As usual in the literature, we will only consider models that assume a   spherically symmetric density profile, characterized by a same value for the radius and mass of the Earth. From seismic studies, we know that the Earth consists of concentric shells of different densities and compositions, with the detailed distribution of density being available in the Preliminary Reference Earth Model (PREM) \cite{Dziewonski:1981xy}. Such model will provide the baseline predictions in our analysis. A simplified description, which captures the main results from seismic studies that indicate that Earth consists of a crust, a mantle and a core, is given by a model characterized by a three - layered profile. Such a model will be denoted by core - mantle - crust. We will also consider two alternative models for a two - layered profile, which can be obtained from the three layered profile by fusing the  core and mantle  together, or by merging the crust and mantle in a unique layer.
These two - layered models will be denoted mantle - crust and core - mantle profiles hereafter.  Finally, we also consider a one - layer profile, characterized by a uniform density profile inside the Earth.
The boundaries and densities of the different layers present in these distinct models are presented in Table \ref{tab:profiles}. In Fig. \ref{fig:earthProfile} (a) we present the associated density distributions as a function of the radial distance crossed by an incoming neutrino with $\mathrm{cos}\, \theta_z = -1$. One has that the main difference among the models occurs in the transition between the low and high density regimes of the Earth profile, with the PREM predicting the largest density for the center of Earth. The impact of these different models on the column depth $X$, defined in Eq. (\ref{eq:column}), is presented in Fig. \ref{fig:earthProfile} (b) as a function of $\mathrm{cos}\, \theta_z$.
In general, the predictions derived using Core-Mantle-Crust and Core-Mantle models are similar to those using the PREM profile. The Mantle-Crust and Uniform models are the ones that differ most from PREM. In particular, for $\mathrm{cos}\,\theta_z = -1$, PREM describes $X$ as being $42\%$ and $57\%$ greater than these simplified models, respectively. In the next section, we will present our results for the neutrino attenuation derived considering these different models for the density profile.

\begin{table}[t]
    \centering
    \begin{tabular}{|c|c|c|} \hline 
             Model       & Layer Boundaries (km) & Layer densities (g/cm$^3$) \\ \hline \hline
     PREM   & Multi layers         & Multi densities                \\ \hline 
         Core - mantle - crust          & (0,3480,5701,6371)             & (11.37,5,3.3)                     \\ \hline 
         Mantle - crust          & (0,5701,6371)             & (6.45,3.3)                     \\ \hline 
         Core - mantle           & (0,3480,6371)             & (11.37,4.42)                     \\ \hline 
         Uniform          & (0,6371)             & (5.55)                     \\ \hline \hline
         
    \end{tabular}
    \caption{Boundaries and densities of layers associated with the distinct models for the density profile of Earth considered in our analysis.}
    \label{tab:profiles}
    
\end{table}

%\section{Neutrino propagation through the Earth}

\begin{figure}[t]
	\centering
	\begin{tabular}{ccc}
	\includegraphics[width=0.45\textwidth]{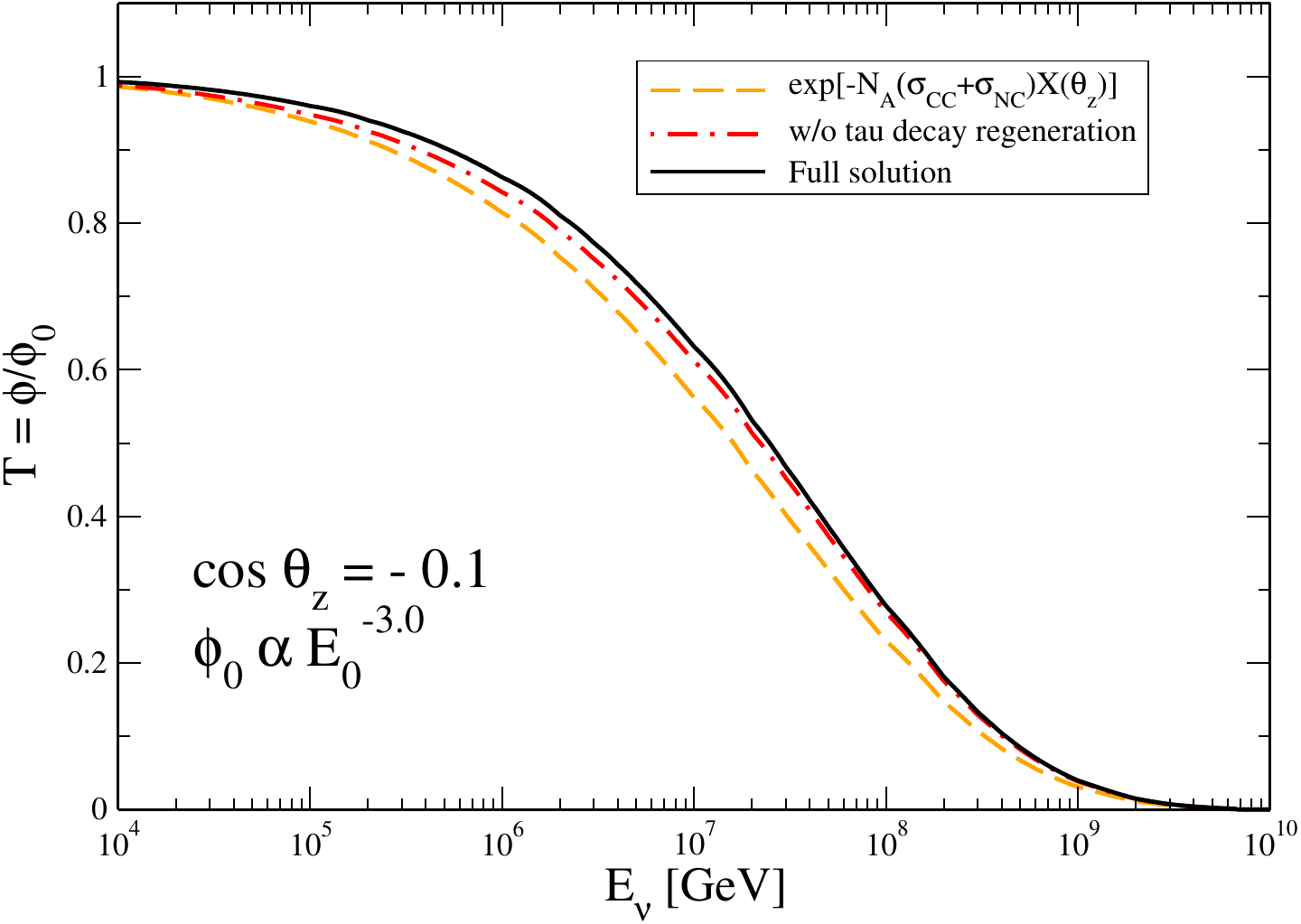} & \,\,\,\,\, & \includegraphics[width=0.45\textwidth]{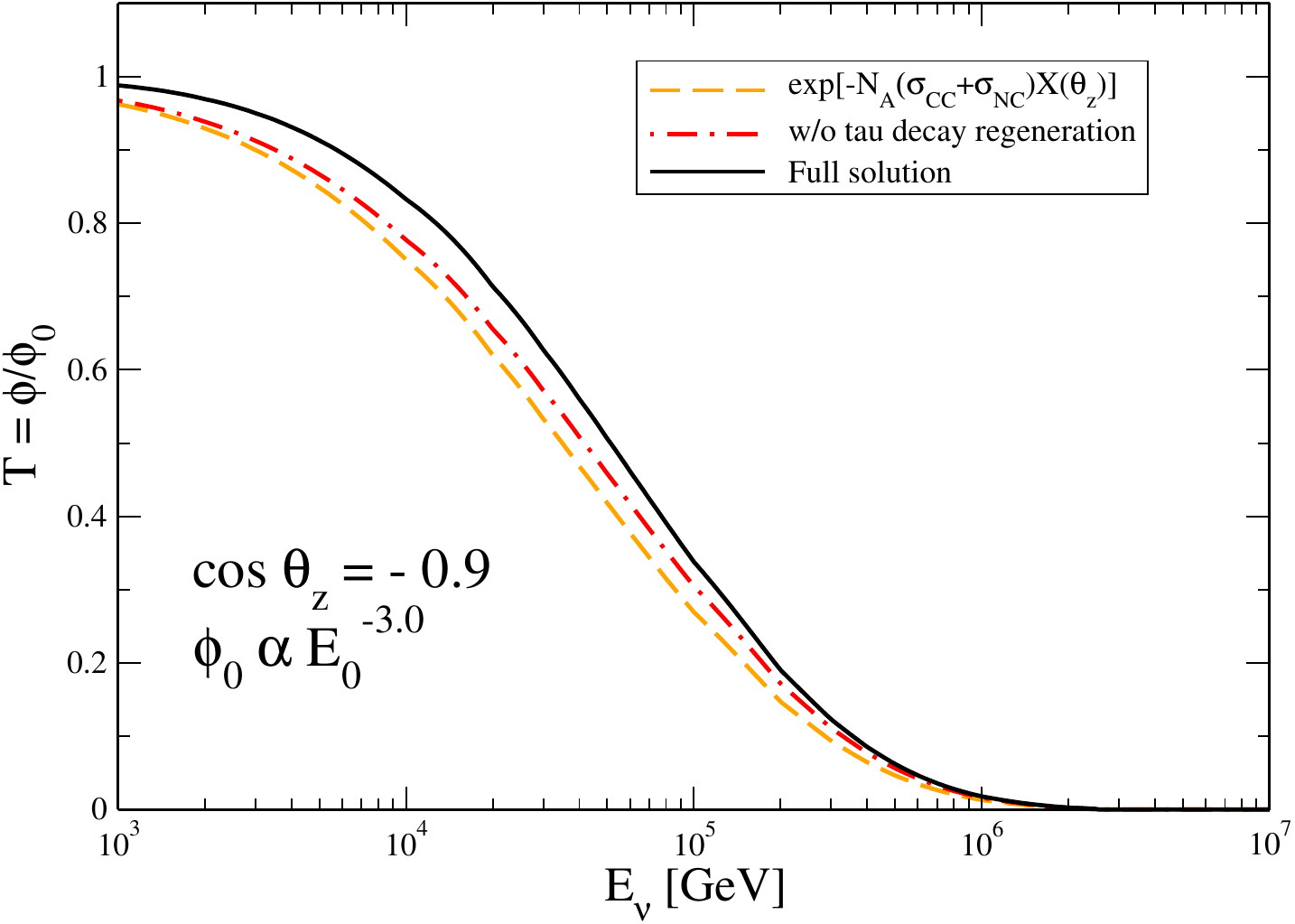} \\
	(a) & \, &  (b) 
			\end{tabular}
\caption{Predictions for the energy dependence of the transmission coefficient for the tau neutrino flux, derived considering different treatments for the calculation of the neutrino propagation. Results for (a) $\mathrm{cos}\, \theta_z = -0.1$ and (b) $\mathrm{cos}\, \theta_z = -0.9$.}
\label{fig:T_comparation}
\end{figure}

\begin{figure}[t]
	\centering
	\begin{tabular}{ccc}
	\includegraphics[width=0.45\textwidth]{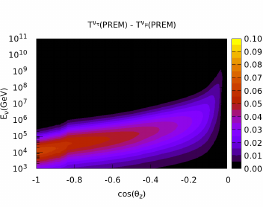} & \,\,\,\,\, & \includegraphics[width=0.45\textwidth]{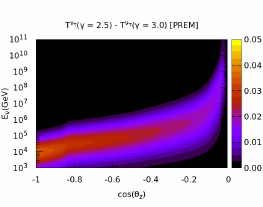} \\
	(a) & \, &  (b) 
			\end{tabular}
\caption{(a)  Difference between the transmission coefficients of the flux of tau and muon neutrinos. (b) Angular and energy dependence of the difference between the transmission coefficient of the tau neutrino flux considering $\gamma = 2.5$ and $\gamma = 3.0$. In both cases, the PREM density profile is assumed. }
\label{fig:diffs}
\end{figure}

\section{Results}
\label{sec:results}

In what follows, we will present our results for the energy and zenith angle dependencies of the transmission coefficient $T$, defined as  the ratio between the surviving and incident neutrino fluxes, considering the different models for the density profile of Earth discussed in the previous section. Initially, we will investigate the impact of the neutral current interactions and flux regeneration  on the tau neutrino transmission coefficient calculated using PREM profile and a spectral index $\gamma = 3$. As discussed in e.g. Ref. \cite{Garcia:2020jwr}, when  the effects of neutral current interaction and flux regeneration by tau decays are neglected,  Eq. (\ref{eq:fluxoNu}) has an analytical solution given by $T = \mathrm{exp}[-N_A \sigma X(\theta_z)]$, with $\sigma = \sigma^{CC} + \sigma^{NC}$. In this case, $T$ can be interpreted as the probability of survival of a neutrino of energy $E_\nu$ incident with angle $\theta_z$. In Fig.  \ref{fig:T_comparation}, we present a comparison between the solution of the cascade equations, given by  Eqs. (\ref{eq:fluxoNu}) and (\ref{eq:fluxoTau}), and  approximated solutions as a function of the neutrino energy $E_\nu$ for (a) $\mathrm{cos}\,\theta_z$ = -0.1  and (b) $\mathrm{cos}\,\theta_z$ = -0.9. As the predictions are strongly dependent on the zenith angle, we have considered different energy ranges in the $x$ - axes of the two plots. One has that $T$ decreases faster with the neutrino energy for  $\mathrm{cos}\,\theta_z$ = -0.9, which is associated with the larger amount of matter crossed by the neutrino when traversing the Earth. 
The comparison between the full  and exponential solutions indicates that the approximated expression  underestimate  the transmission coefficient. Such result is expected, since in the approximated solution we are disregarding two contributions that lead to a smaller decreasing of $T$ with the neutrino energy for a fixed value of $\mathrm{cos}\,\theta_z$. The impact of the tau decay regeneration can be estimated by comparing the full solution with the result derived by neglecting the third term in Eq. (\ref{eq:fluxoNu}), which is represented by the dashed - dotted line in Fig. \ref{fig:T_comparation}.  One has that the contribution of the tau - decay regeneration becomes non - negligible for larges values of $\theta_z$, enhancing the transmission coefficient for smaller neutrino energies, as expected from the discussion presented in the previous section. 

In Fig. \ref{fig:diffs} (a) we present our predictions for the difference between the transmission coefficients associated with the tau and muon neutrino fluxes. One has that the difference is ever larger than zero, due to the tau regeneration process. The differences are greater for neutrinos that cross through a larger amount of  matter. In particular, for neutrinos that cross the Earth's core ($\mathrm{cos}\,\theta_z \leq -0.84$) with energy of $10^4\,\mathrm{GeV}$ the differences reach a value of 0.064.  Such difference decreases to 0.021 for $E_\nu \approx 10^5\,\mathrm{GeV}$. 
  One has verified that the difference is more significant for neutrinos than for antineutrinos, which is  related with the distribution $dn/dy$ present in Eq. (\ref{eq:fluxoNu}), which predicts that the neutrino generated in the tau decay carries a larger amount of the tau energy. 
  
  The previous results were derived for a spectral index $\gamma = 3$.
However, IceCube measurements have pointed out for a different value for $\gamma$ in the range 2.2 -- 2.87, depending on the events selected in the analysis \cite{HESE1,HESE2,tracksTotal,tracksNorte,cascades}.  In order to estimate the impact of a different spectral index in our predictions for the transmission coefficient of the tau neutrino flux, we have solved the cascade equations for $\gamma = 2.5$ and in Fig. \ref{fig:diffs} (b) we present the results for the difference between the predictions for the two spectral indexes, derived assuming the PREM profile. One has that the difference in spectral index becomes more important for the region when the neutrino cross through a greater amount of matter. For $\mathrm{cos}\,\theta_z = -1$ the transmission coefficient  increases by a value of 0.016, 0.036 and 0.008 for energies of $10^3$, $10^4$ and $10^5\,\mathrm{GeV}$, respectively, when we decrease the spectral index from 3.0 to 2.5.
In what follows, we will only present the results for $\gamma = 3.0$, but the predictions for other values of $\gamma$ are available upon request.

%For example,  the first analysis of the High Energy Start Events (HESE) obtained $\gamma = 2.2$ \cite{HESE1}. In recent years, IceCube has updated this parameter with new data collected, and also estimated it with different types of event topologies seen at the observatory. The HESE data accumulated in 7.5 years implies $\gamma = 2.87$ \cite{HESE2}. With analysis only of tracks for the entire sky in 10.3 years, $\gamma = 2.58$ \cite{tracksTotal} was obtained, whereas for tracks only coming from the northern hemisphere in 12.3 years, $\gamma = 2.38$ \cite{tracksNorte} was estimated. Finally, in analyzes of all-sky cascades only, the most recent results indicate $\gamma = 2.53$ \cite{cascades}.

\begin{figure}[t]
	\centering
	\begin{tabular}{ccc}
    \includegraphics[width=0.33\textwidth]{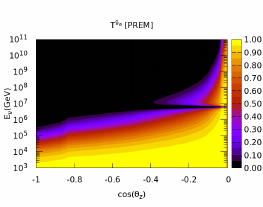} &\includegraphics[width=0.33\textwidth]{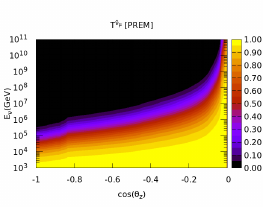} & \includegraphics[width=0.33\textwidth]{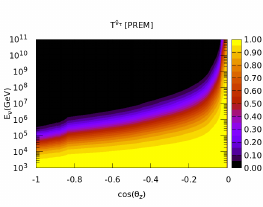} \\
    \includegraphics[width=0.33\textwidth]{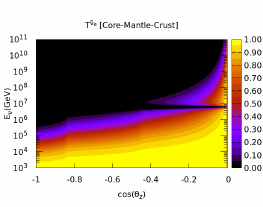} &\includegraphics[width=0.33\textwidth]{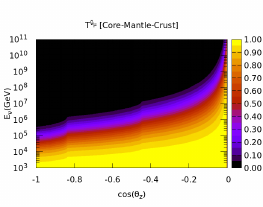} & \includegraphics[width=0.33\textwidth]{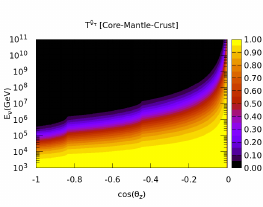} \\
    \includegraphics[width=0.33\textwidth]{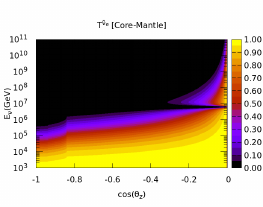} & \includegraphics[width=0.33\textwidth]{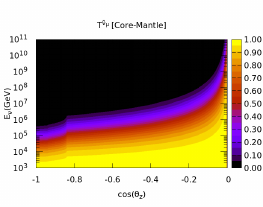} & \includegraphics[width=0.33\textwidth]{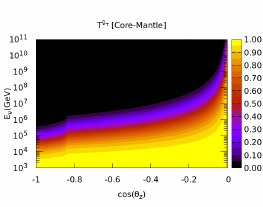} \\
    \includegraphics[width=0.33\textwidth]{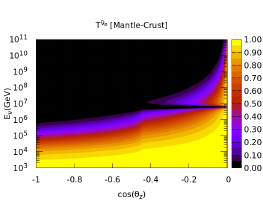} & \includegraphics[width=0.33\textwidth]{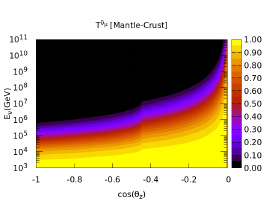} & \includegraphics[width=0.33\textwidth]{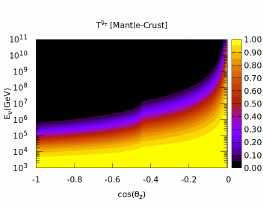} \\
    \includegraphics[width=0.33\textwidth]{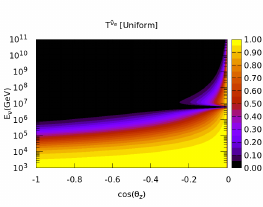} & \includegraphics[width=0.33\textwidth]{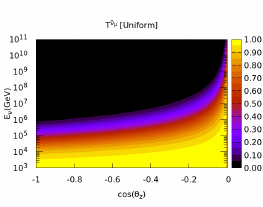} & \includegraphics[width=0.33\textwidth]{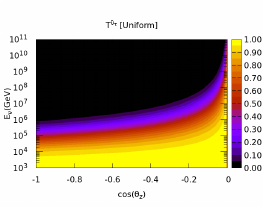} %\\
%	(a) & \, &  (b) 
			\end{tabular}
\caption{ Angular and energy dependencies of transmission coefficient for  electron (left), muon (middle), and tau (right) antineutrinos crossing the Earth. The predictions derived using the distinct models for the density profile are presented in different lines (see title of the plots). }
\label{fig:transmissionAnu}
\end{figure}

\begin{figure}[t]
	\centering
	\begin{tabular}{ccc}
	\includegraphics[width=0.4\textwidth]{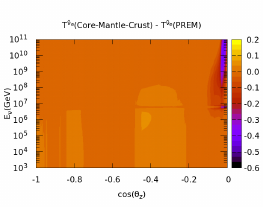} & \,\,\,\,\, & \includegraphics[width=0.4\textwidth]{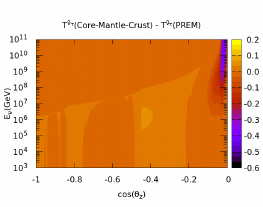} \\
	\includegraphics[width=0.4\textwidth]{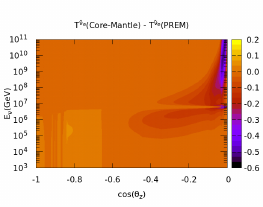} & \,\,\,\,\, & \includegraphics[width=0.4\textwidth]{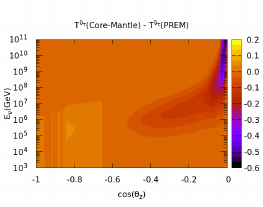} \\
	\includegraphics[width=0.4\textwidth]{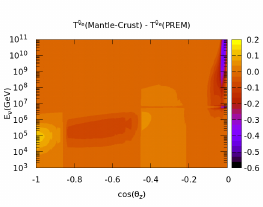} & \,\,\,\,\, & \includegraphics[width=0.4\textwidth]{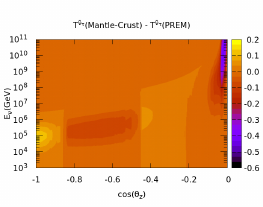} \\
	\includegraphics[width=0.4\textwidth]{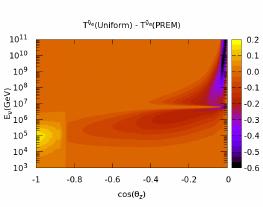} & \,\,\,\,\, & \includegraphics[width=0.4\textwidth]{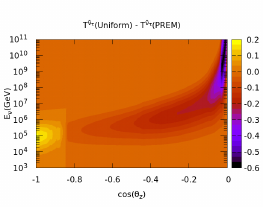} %\\
%	(a) & \, &  (b) 
			\end{tabular}
\caption{Predictions for the difference between the transmission coefficients derived using simplified models for Earth's density profile  and using PREM. Results for electron (left) and tau (right) antineutrinos. }
\label{fig:diffModelsPREM}
\end{figure}

In Fig. \ref{fig:transmissionAnu} our predictions for the transmission coefficient for distinct antineutrino flavors as a function of the initial energy and incidence angle, derived considering the different models for the  density profile \footnote{The results for neutrinos are available from the authors upon a reasonable request.}. The results are organized as follows: (a) the plots presented in a given line represent a specific density profile model, and (b) the different columns present the results for distinct antineutrino flavors, with the predictions for the electron, muon and tau antineutrinos shown in the left, middle and right panels, respectively. The impact of the Glashow resonance is clear on the predictions for the electron antineutrinos, implying that $T \rightarrow 0$ for $E_\nu \approx 6.3\, \mathrm{PeV}$ and all zenith angles, independently of the density profile model considered. Moreover,  the results are not sensitive to the model  for $\rho(r)$ when $\cos{\theta_{z}}\rightarrow -1$, since in this limit  the Earth is opaque to the ultra-high - energy neutrinos. The description of the Earth profile in terms of  few layers has  direct impact on the angular dependence of $T$. In particular, the boundary between layers implies a ''kink" in the angular distribution \cite{Gonzalez-Garcia:2007wfs}, which are visible in the predictions associated  with the   Core-Mantle-Crust, Core-Mantle and Mantle-Crust models. In the PREM case, which assumes the presence of multi layers, this ''kink" is only present for $\cos{\theta_{z}}\rightarrow -0.84$. As expected, such behavior does not occur if we consider the uniform model. The difference between the predictions of the different models and the PREM results for the distinct values of the neutrino energy and zenith angle is quantified in Fig. \ref{fig:diffModelsPREM}. We calculate this difference for electron (left panels) and tau  antineutrinos (right panels). Our results indicate that depending on the model considered, the transmission coefficient can be enhanced or suppressed in a given combination of energy and zenith angle. One has that the simplified models can be a good estimate of neutrino transmission coefficient depending on the energetic and angular region analyzed. In particular, the Core-Mantle-Crust model predicts similar results to those derived using the PREM profile, differing only in the region where $\mathrm{cos}\,\theta_z \rightarrow 0$, that is, for neutrinos that cross a thin layer of the Earth's crust. The Core-Mantle model, like the Core-Mantle-Crust, is similar to PREM for  $\mathrm{cos}\,\theta_z \leq -0.5$, but presents differences of the same order of magnitude as the transmission outside this angular region. In contrast, the Mantle-Crust and Uniform models predict very distinct values for $T$ in comparison with the PREM results.   For these two models, the transmission becomes similar to PREM only in the vicinity of the points where the target column density $X$, shown in Fig. \ref{fig:earthProfile} (b), coincides, i.e., for large values of  $\mathrm{cos}\,\theta_z$.  { The difference between the PREM predictions are those from simplified models are maximized in all comparisons for $\mathrm{cos}\,\theta_z \rightarrow 0$ and ultra-high - energies. This behavior is associated with the greater difference between densities in the Earth's crust, especially closer to the surface, which is the region crossed  by Earth - skimming neutrinos that are probed by ANITA and in forthcoming years by e.g.  Trinity and POEMMA \cite{Ackermann:2022rqc}.}

%On the left are the results for electronic antineutrinos, in the middle muonic ones, and on the right for tauonic ones. The main differences for the transmission of the different flavors reflect the presence of the Glashow resonance for electronic antineutrino interactions ($E_\nu \approx 6.3\, \mathrm{PeV}$), and the regeneration of the tau antineutrino flux by antitau decay, leaving the transmission slightly larger. We are considering an incident neutrino flux as a power law ($\Phi \sim E_\nu ^{-\gamma}$) with spectral index $\gamma = 3.0$. The results for neutrinos have similar behavior, but without the effect of the Glashow resonance and with the magnitude slightly affected by differences in the neutrino-nucleon cross section. In each line of Fig. \ref{fig:transmissionAnu} we use a different model for the Earth's profile density. In the top line we use PREM, a model widely used in similar works. In the remaining lines we use simplified models of three-, two- and one- layer for the Earth, Core-Mantle-Crust, Core-Mantle and Mantle-Crust, and Uniform, respectively.

%In Fig. \ref{fig:diffModelsPREM} we show the difference between the transmission coefficient obtained using the simplified models for the Earth's structure and using PREM. 

\section{Summary}
\label{sec:conc}
The events observed in UHE neutrino detectors  are  strongly dependent  on the neutrino fluxes incident at the Earth and their absorption during the passage through Earth to the detector. The attenuation of the incident neutrino flux depends on the neutrino energy and the arrival direction, with the neutrino propagation depending on the details of the matter structure between the source and the detector. One has that for relatively small values of the neutrino energy ($E_{\nu} \lesssim 50$ TeV), the Earth is essentially transparent to neutrinos, while above it, the neutrinos traveling through a sufficient chord length inside the Earth may interact before  arriving at the detector. The description of this absorption is strongly dependent on  the total amount of matter that neutrino feel as a function of zenith angle $\theta_z$. In this paper, we have investigated the impact of different descriptions for the density profile of Earth on the transmission coefficient. In particular, we have considered five different models, being one of them the PREM profile, which characterizes the current state of the art. We have estimated how these different models modify the target column density and estimated the transmission coefficient for different flavors by solving the cascade equations taking into account the NC interactions and tau regeneration. A comparison with approximated solutions is also performed. Our results indicated that the predictions are sensitive to the model considered for the density profile, with the simplified three layer model providing a satisfactory description when compared with the PREM results. In contrast, models that does not take into account the presence of a core, implying very distinct results for the transmission coefficient.

\begin{acknowledgments}

R. F. acknowledges support from the Conselho Nacional de Desenvolvimento Cient\'{\i}fico e Tecnol\'ogico (CNPq, Brazil), Grant No. 161770/2022-3. D.R.G. V.P.G. was partially supported by CNPq, FAPERGS and INCT-FNA (Process No. 464898/2014-5). D.R.G.  was partially supported by CNPq.

\end{acknowledgments}

%\section*{Declaration}

%{\bf Autor contribution} R.F. and D.R.G. made the formal analysis and code implementation in the software. V.P.G. wrote the main manuscript text and managed the project. All authors reviewed the manuscript.

%\hspace{0.5cm}

%{\bf Ethics declaration} Not applicable.

%\hspace{0.5cm}

%{\bf Data Availability} No datasets were generated or analyzed during the
%current study.

%\hspace{0.5cm}

%{\bf Competing Interests} The authors declare no competing interests.

\hspace{1.0cm}


\begin{thebibliography}{99}

\bibitem{Ackermann:2022rqc}
M.~Ackermann, M.~Bustamante, L.~Lu, N.~Otte, M.~H.~Reno, S.~Wissel, S.~K.~Agarwalla, J.~Alvarez-Mu\~niz, R.~Alves Batista and C.~A.~Arg\"uelles, \textit{et al.}
%``High-energy and ultra-high-energy neutrinos: A Snowmass white paper,''
JHEAp \textbf{36}, 55-110 (2022)


\bibitem{MammenAbraham:2022xoc}
R.~Mammen Abraham, J.~Alvarez-Mu\~niz, C.~A.~Arg\"uelles, A.~Ariga, T.~Ariga, A.~Aurisano, D.~Autiero, M.~Bishai, N.~Bostan and M.~Bustamante, \textit{et al.}
%``Tau neutrinos in the next decade: from GeV to EeV,''
J. Phys. G \textbf{49}, no.11, 110501 (2022)

\bibitem{Reno:2023sdm}
M.~H.~Reno,
%``High-Energy to Ultrahigh-Energy Neutrino Interactions,''
Ann. Rev. Nucl. Part. Sci. \textbf{73}, no.1, 181-204 (2023)


\bibitem{CS}
A.~Cooper-Sarkar, P.~Mertsch and S.~Sarkar,
%``The high energy neutrino cross-section in the Standard Model and its uncertainty,''
JHEP \textbf{08}, 042 (2011)

\bibitem{Goncalves:2010ay}
V.~P.~Goncalves and P.~Hepp,
%``A comparative study of the neutrino-nucleon cross section at ultra high energies,''
Phys. Rev. D \textbf{83}, 014014 (2011)

\bibitem{Illarionov:2011wc}
A.~Y.~Illarionov, B.~A.~Kniehl and A.~V.~Kotikov,
%``Ultrahigh-energy neutrino-nucleon deep-inelastic scattering and the Froissart bound,''
Phys. Rev. Lett. \textbf{106} (2011), 231802



\bibitem{Block:2013nia}
M.~M.~Block, L.~Durand, P.~Ha and D.~W.~McKay,
%``Implications of a Froissart bound saturation of \ensuremath{\gamma}*-p deep inelastic scattering. II. Ultrahigh energy neutrino interactions,''
Phys. Rev. D \textbf{88}, no.1, 013003 (2013)





\bibitem{Goncalves:2015fua}
V.~P.~Goncalves and D.~R.~Gratieri,
%``Investigating the effects of the QCD dynamics in the neutrino absorption by the Earth\textquoteright{}s interior at ultrahigh energies,''
Phys. Rev. D \textbf{92}, no.11, 113007 (2015)


\bibitem{Albacete:2015zra}
J.~L.~Albacete, J.~I.~Illana and A.~Soto-Ontoso,
%``Neutrino-nucleon cross section at ultrahigh energy and its astrophysical implications,''
Phys. Rev. D \textbf{92}, no.1, 014027 (2015)


\bibitem{Arguelles:2015wba}
C.~A.~Arg\"uelles, F.~Halzen, L.~Wille, M.~Kroll and M.~H.~Reno,
%``High-energy behavior of photon, neutrino, and proton cross sections,''
Phys. Rev. D \textbf{92}, no.7, 074040 (2015)


\bibitem{bgr18}
V.~Bertone, R.~Gauld and J.~Rojo,
%``Neutrino Telescopes as QCD Microscopes,''
JHEP \textbf{01}, 217 (2019)


%\cite{Goncalves:2021gcu}
\bibitem{Goncalves:2021gcu}
V.~P.~Gon\c{c}alves, D.~R.~Gratieri and A.~S.~C.~Quadros,
%``Estimating the impact of the QCD dynamics on the determination of the high energy astrophysical neutrino flux,''
Eur. Phys. J. C \textbf{81}, no.6, 496 (2021)






\bibitem{Valera:2022ylt}
V.~B.~Valera, M.~Bustamante and C.~Glaser,
%``The ultra-high-energy neutrino-nucleon cross section: measurement forecasts for an era of cosmic EeV-neutrino discovery,''
JHEP \textbf{06}, 105 (2022)

\bibitem{Goncalves:2022uwy}
V.~P.~Goncalves, D.~R.~Gratieri and A.~S.~C.~Quadros,
%``Implications of the QCD dynamics and a Super-Glashow astrophysical neutrino flux on the description of ultrahigh energy neutrino data,''
Eur. Phys. J. C \textbf{82}, no.11, 1011 (2022)


  \bibitem{hdqcd} 
  F.~Gelis, E.~Iancu, J.~Jalilian-Marian and R.~Venugopalan,
    Ann.\ Rev.\ Nucl.\ Part.\ Sci.\  {\bf 60}, 463 (2010);
  H.~Weigert,  Prog.\ Part.\ Nucl.\ Phys.\  {\bf 55}, 461 (2005); J.~Jalilian-Marian and Y.~V.~Kovchegov, Prog.\ Part.\ Nucl.\ Phys.\  {\bf 56}, 104 (2006).


\bibitem{Fiorillo:2022rft}
D.~F.~G.~Fiorillo and M.~Bustamante,
%``Bump hunting in the diffuse flux of high-energy cosmic neutrinos,''
Phys. Rev. D \textbf{107}, no.8, 083008 (2023)


\bibitem{Dziewonski:1981xy}
A.~M.~Dziewonski and D.~L.~Anderson,
%``Preliminary reference earth model,''
Phys. Earth Planet. Interiors \textbf{25}, 297-356 (1981)


\bibitem{bolt}
  B. A. Bolt, Q. J. R. Astron. Soc. \textbf{32}, 367 (1991).

\bibitem{kennett}
B. L. N. Kennett, Geophys. J. Int. \textbf{132}, 374 (1998).


\bibitem{masters}
G. Masters and D. Gubbins, Phys. Earth Planet. Inter. \textbf{140}, 159 (2003).

\bibitem{wit}
R. de Wit, P. K\"aufl, A. Valentine, and J. Trampert, Phys. Earth Planet. Inter. \textbf{237}, 1 (2014).






\bibitem{Volkova:1974xa}
L.~V.~Volkova and G.~T.~Zatsepin,
%``On the problem of neutrino penetration though the earth. (talk, in russian),''
Izv. Akad. Nauk Ser. Fiz. \textbf{38N5}, 1060-1063 (1974)
%5 citations counted in INSPIRE as of 19 Mar 2024


\bibitem{DeRujula:1983ya}
A.~De Rujula, S.~L.~Glashow, R.~R.~Wilson and G.~Charpak,
%``Neutrino Exploration of the Earth,''
Phys. Rept. \textbf{99}, 341 (1983)


\bibitem{Wilson:1983an}
T.~L.~Wilson,
%``Neutrino Tomography: Tevatron Mapping Versus the Neutrino Sky,''
Nature \textbf{309}, 38-42 (1984)


\bibitem{Jain:1999kp}
P.~Jain, J.~P.~Ralston and G.~M.~Frichter,
%``Neutrino absorption tomography of the earth's interior using isotropic ultrahigh-energy flux,''
Astropart. Phys. \textbf{12}, 193-198 (1999)



\bibitem{Reynoso:2004dt}
M.~M.~Reynoso and O.~A.~Sampayo,
%``On neutrino absorption tomography of the earth,''
Astropart. Phys. \textbf{21}, 315-324 (2004)

\bibitem{Gonzalez-Garcia:2007wfs}
M.~C.~Gonzalez-Garcia, F.~Halzen, M.~Maltoni and H.~K.~M.~Tanaka,
%``Radiography of earth's core and mantle with atmospheric neutrinos,''
Phys. Rev. Lett. \textbf{100}, 061802 (2008)


\bibitem{Borriello:2009ad}
E.~Borriello, G.~Mangano, A.~Marotta, G.~Miele, P.~Migliozzi, C.~A.~Moura, S.~Pastor, O.~Pisanti and P.~E.~Strolin,
%``Sensitivity on Earth Core and Mantle densities using Atmospheric Neutrinos,''
JCAP \textbf{06}, 030 (2009)

\bibitem{Denton:2020jft}
P.~B.~Denton and Y.~Kini,
%``Ultra-High-Energy Tau Neutrino Cross Sections with GRAND and POEMMA,''
Phys. Rev. D \textbf{102}, 123019 (2020)

\bibitem{Kumar:2021faw}
A.~Kumar and S.~K.~Agarwalla,
%``Validating the Earth\textquoteright{}s core using atmospheric neutrinos with ICAL at INO,''
JHEP \textbf{08}, 139 (2021)

\bibitem{Hajjar:2023knk}
R.~Hajjar, O.~Mena and S.~Palomares-Ruiz,
%``Earth tomography with supernova neutrinos at future neutrino detectors,''
Phys. Rev. D \textbf{108}, no.8, 083011 (2023)


\bibitem{Donini:2018tsg}
A.~Donini, S.~Palomares-Ruiz and J.~Salvado,
%``Neutrino tomography of Earth,''
Nature Phys. \textbf{15}, no.1, 37-40 (2019)



\bibitem{Denton:2021rgt}
P.~B.~Denton and R.~Pestes,
%``Neutrino oscillations through the Earth\textquoteright{}s core,''
Phys. Rev. D \textbf{104}, no.11, 113007 (2021)


\bibitem{Wolfenstein}
 L. Wolfenstein, Phys. ReV. D 17, 2369 (1978)


\bibitem{MS}
S.~P.~Mikheyev and A.~Y.~Smirnov,
%``Resonance Amplification of Oscillations in Matter and Spectroscopy of Solar Neutrinos,''
Sov. J. Nucl. Phys. \textbf{42}, 913-917 (1985)
%4150 citations counted in INSPIRE as of 08 May 2024


%\cite{Kelly:2021jfs}
\bibitem{Kelly:2021jfs}
K.~J.~Kelly, P.~A.~N.~Machado, I.~Martinez-Soler and Y.~F.~Perez-Gonzalez,
%``DUNE atmospheric neutrinos: Earth tomography,''
JHEP \textbf{05}, 187 (2022)
%doi:10.1007/JHEP05(2022)187
%[arXiv:2110.00003 [hep-ph]].
%23 citations counted in INSPIRE as of 07 May 2024

%\cite{Gratieri:2017anb}
\bibitem{Gratieri:2017anb}
D.~R.~Gratieri, M.~M.~Guzzo and O.~L.~G.~Peres,
%``Impact of standard neutrino oscillations and systematics on proton lifetime measurements,''
J. Phys. G \textbf{46}, no.7, 075006 (2019)
%doi:10.1088/1361-6471/ab0b56
%[arXiv:1704.03927 [hep-ph]].
%2 citations counted in INSPIRE as of 07 May 2024






  

 \bibitem{book}
M.~S.~Athar and S.~K.~Singh,
{\it The Physics of Neutrino Interactions,}
Cambridge University Press, 2020.



\bibitem{Niess:2018opy}
V.~Niess and O.~Martineau-Huynh,
%``DANTON: a Monte-Carlo sampler of $\tau$ from $\nu_\tau$ interacting with the Earth,''
[arXiv:1810.01978 [physics.comp-ph]].




\bibitem{Safa:2021ghs}
I.~Safa, J.~Lazar, A.~Pizzuto, O.~Vasquez, C.~A.~Arg\"uelles and J.~Vandenbroucke,
%``TauRunner: A public Python program to propagate neutral and charged leptons,''
Comput. Phys. Commun. \textbf{278}, 108422 (2022)



\bibitem{Garg:2022ugd}
D.~Garg, S.~Patel, M.~H.~Reno, A.~Reustle, Y.~Akaike, L.~A.~Anchordoqui, D.~R.~Bergman, I.~Buckland, A.~L.~Cummings and J.~Eser, \textit{et al.}
%``Neutrino propagation in the Earth and emerging charged leptons with nuPyProp,''
JCAP \textbf{01}, 041 (2023)

\bibitem{Garcia:2020jwr}
A.~Garcia, R.~Gauld, A.~Heijboer and J.~Rojo,
%``Complete predictions for high-energy neutrino propagation in matter,''
JCAP \textbf{09}, 025 (2020)



  \bibitem{edo1}
A.~Nicolaidis and A.~Taramopoulos,
  %``Shadowing of ultrahigh energy neutrinos,''
 Phys.\ Lett.\ B {\bf 386}, 211 (1996)
 
  \bibitem{edo2}
V.~A.~Naumov and L.~Perrone,
  %``Neutrino propagation through dense matter,''
 Astropart.\ Phys. {\bf 10}, 239 (1999)

 \bibitem{edo3}
J.~Kwiecinski, A.~D.~Martin, and A.~M.~Stasto,
  %``Penetration of the Earth by ultrahigh energy neutrinos predicted by low x QCD,''
 Phys.\ Rev.\ D {\bf 59}, 093002 (1999)

  \bibitem{edo4}
S.~Iyer, M.~H.~Reno, and I.~Sarcevic,
  %``Searching for νμ→ντ oscillations with extragalactic neutrinos,''
 Phys.\ Rev.\ D {\bf 61}, 053003 (2000)

  \bibitem{edo5}
K.~Giesel, J.~H.~Jureit, and E.~Reya,
  %``Cosmic UHE neutrino signatures,''
 Astropart.\ Phys. {\bf 20}, 335 (2003)

  \bibitem{edo6}
E.~Reya and J.~Rodiger,
  %``Signatures of cosmic tau neutrinos,''
 Phys.\ Rev.\ D {\bf 72}, 053004 (2005)

   \bibitem{edo7}
S.~Rakshit and E.~Reya,
  %``Transport equations of cosmic neutrinos passing through Earth and secondary νμ fluxes,''
 Phys.\ Rev.\ D {\bf 74}, 103006 (2006)

    \bibitem{edo8}
S.~Palomares-Ruiz, A.~C.~Vincent, and O.~Mena,
  %``Spectral analysis of the high-energy IceCube neutrinos,''
 Phys.\ Rev.\ D {\bf 91}, 103008 (2015)

\bibitem{Alvarez-Muniz:2018owm}
J.~Alvarez-Mu\~niz, W.~R.~Carvalho, A.~L.~Cummings, K.~Payet, A.~Romero-Wolf, H.~Schoorlemmer and E.~Zas,
%``Comprehensive approach to tau-lepton production by high-energy tau neutrinos propagating through the Earth,''
Phys. Rev. D \textbf{97}, no.2, 023021 (2018)
[erratum: Phys. Rev. D \textbf{99}, no.6, 069902 (2019)]



    \bibitem{paramet1}
S.~I.~Dutta, M.~H.~Reno, and I.~Sarcevic,
  %``Tau neutrinos underground: Signals of νμ→ντ oscillations with extragalactic neutrinos,''
 Phys.\ Rev.\ D {\bf 62}, 123001 (2000).


 \bibitem{paschos}
E.~A.~Paschos and J.~Y.~Yu,
  %``Neutrino interactions in oscillation experiments,''
 Phys.\ Rev.\ D {\bf 65}, 033002 (2002)

 \bibitem{reno}
S.~Kretzer and M.~H.~Reno,
  %``Tau neutrino deep inelastic charged current interactions,''
 Phys.\ Rev.\ D {\bf 66}, 113007 (2002)

     
 \bibitem{ct14} S.~Dulat {\it et al.}, Phys. Rev. D {\bf 93}, 033006 (2016).


%%%%%%%%%%%%%%%%%%%%%%%%%%%%%%%%%%%%%%%%%%%%%%%%%%%




     \bibitem{HESE1}
M.~G.~Aartsen {\it et al.} [IceCube Collaboration],
  %``Evidence for High-Energy Extraterrestrial Neutrinos at the IceCube Detector,''
 Science {\bf 342}, 1242856 (2013)

     \bibitem{HESE2}
R.~Abbasi {\it et al.} [IceCube Collaboration],
  %``The IceCube high-energy starting event sample: Description and flux characterization with 7.5 years of data,''
 Phys.\ Rev.\ D {\bf 104}, 022002 (2021)

      \bibitem{tracksTotal}
R.~Abbasi {\it et al.} [IceCube Collaboration],
  %``Characterization of the Astrophysical Diffuse Neutrino Flux using Starting Track Events in IceCube,''
 [arXiv:2402.18026 [astro-ph.HE]]

       \bibitem{tracksNorte}
P.~Fuerst {\it et al.} [IceCube Collaboration],
  %``Galactic and Extragalactic Analysis of the Astrophysical Muon Neutrino Flux with 12.3 years of IceCube Track Data,''
 Pos {\bf ICRC2023} 1046 (2023)

     \bibitem{cascades}
M.~G.~Aartsen {\it et al.} [IceCube Collaboration],
  %``Characteristics of the Diffuse Astrophysical Electron and Tau Neutrino Flux with Six Years of IceCube High Energy Cascade Data,''
 Phys.\ Rev.\ Lett. {\bf 125}, 121104 (2020)

\end{thebibliography}
\end{document}